\documentclass[useAMS,usenatbib]{mn2e}

\usepackage{epsfig}
\usepackage{amsmath}
\usepackage{amssymb}
\usepackage{natbib}
\usepackage{multirow}
\usepackage{graphicx} 
\usepackage{color}
\usepackage{epstopdf}

\newcommand{\eprint}[2][]{{\tt\if!#1!#2\else#1:#2\fi}}

\title[Co-Evolution of Total Density Profiles and Central Dark Matter Fractions]{The Co-Evolution of Total Density Profiles and Central Dark Matter Fractions in Simulated Early-Type Galaxies}
\author[Remus et al.]{Rhea-Silvia Remus$^{1}$\thanks{E-mail:\textit{rhea@usm.lmu.de}}, Klaus Dolag$^{1,2}$, Thorsten Naab$^{2}$, Andreas Burkert$^{1,3}$,
\newauthor Michaela Hirschmann$^{4}$, Tadziu L. Hoffmann$^{1}$, Peter H. Johansson$^{5}$\\
$^{1}$ Universit\"ats-Sternwarte M\"unchen, Scheinerstr.\ 1, D-81679 M\"unchen, Germany\\
$^{2}$ Max Planck Institut for Astrophysics, D-85748 Garching, Germany\\
$^{3}$ Max Planck Institut for Extraterrestrial Physics, D-85748 Garching, Germany\\
$^{4}$ Sorbonne Universites, UPMC-CNRS, UMR7095, Institut d'Astrophysique de Paris, F-75014, Paris, France\\
$^{5}$ Department of Physics, University of Helsinki, Gustaf H\"allstr\"omin katu 2a, FI-00014 Helsinki, Finland}
\begin{document}

\date{Accepted ... Received ...; in original form ...}

\pagerange{\pageref{firstpage}--\pageref{lastpage}} \pubyear{2015}

\maketitle

\label{firstpage}

\begin{abstract}
We present evidence from cosmological hydrodynamical simulations for a co-evolution of the slope of the total (dark and stellar) mass density profile, $\gamma_\mathrm{tot}$, and the dark matter fraction within the half-mass radius, $f_\mathrm{DM}$, in early-type galaxies.
The relation can be described as $\gamma_\mathrm{tot}=A\,f_\mathrm{DM}+B$ for all systems at all redshifts.
The trend is set by the decreasing importance of gas dissipation towards lower redshifts and for more massive systems.
Early-type galaxies are smaller, more concentrated, have lower $f_\mathrm{DM}$ and steeper $\gamma_\mathrm{tot}$ at high redshifts and at lower masses for a given redshift; $f_\mathrm{DM}$ and $\gamma_\mathrm{tot}$ are good indicators for growth by ``dry'' merging.
The values for $A$ and $B$ change distinctively for different feedback models, and this relation can be used as a test for such models.
A similar correlation exists between $\gamma_\mathrm{tot}$ and the stellar mass surface density $\Sigma_*$.
A model with weak stellar feedback and feedback from black holes is in best agreement with observations.
All simulations, independent of the assumed feedback model, predict steeper $\gamma_\mathrm{tot}$ and lower $f_\mathrm{DM}$ at higher redshifts.
While the latter is in agreement with the observed trends, the former is in conflict with lensing observations, which indicate constant or decreasing $\gamma_\mathrm{tot}$.
This discrepancy is shown to be artificial: the observed trends can be reproduced from the simulations using observational methodology to calculate the total density slopes.
\end{abstract}

\begin{keywords}
galaxies: evolution -- formation -- cosmology: dark matter -- methods: numerical
\vspace*{-6pt}
\end{keywords}

\section{Introduction}
The distribution of dark and luminous matter in early-type galaxies (ETGs) has been the subject of many studies over the past years.
The total mass of galaxies can, for example, be estimated from the X-ray emission of hot halo gas \citep[e.g.,][]{pratt:2005,pointecouteau:2005,das:2011,newman:2013}, strong lensing measurements \citep[e.g.,][]{koopmans:2006,auger:2010ApJ...724..511A,barnabe:2011MNRAS.415.2215B}, or dynamical modelling at different levels of complexity \citep[e.g.,][]{gerhard:2001,thomas:2007MNRAS.382..657T,cappellari:2013b,tortora:2014b}.
More recently, measurements using tracer populations such as planetary nebulae or globular clusters have also been able to assess mass distributions and stellar kinematics out to large radii \citep[e.g.,][]{deason:2012ApJ...748....2D,napolitano:2014,cappellari:2015b}.
The contribution of the dark matter is typically estimated by subtracting the luminous stellar component.
Uncertainties in the IMF slope and thus the mass-to-light ratio, however, can result in significant variations in the derived dark matter fractions \citep[e.g.,][]{barnabe:2011MNRAS.415.2215B,dutton:2013,deason:2012ApJ...748....2D}, i.e., a bottom-heavy IMF, for example, results in more stellar mass for a given stellar luminosity, lowering the dark matter fraction.

Dynamical modelling of Coma cluster ETGs \citep{thomas:2007MNRAS.382..657T} as well as numerical simulations \citep{remus:2013} indicate that the total (stellar plus dark matter) radial density profiles of ETGs at $z=0$ can be well described by a power law $\rho \propto r^\gamma$, with an average slope of $\gamma \approx -2.1$ over a large radial range of $0.3\,R_{1/2}<r<4\,R_{1/2}$.
This is also in good agreement with the results presented by \citet{humphrey:2010}, who studied 10 early-type galaxies (ETGs) in different environments from field to cluster densities at low redshifts. 
They modeled the total mass profiles of ETGs using X-ray measurements of the hot halo surrounding the galaxies, covering $0.2\,R_\mathrm{eff}<r<10\,R_\mathrm{eff}$, and find power-law slopes for the total density between $-2 <\gamma < -1.2$.
Larger galaxies tend to have flatter slopes, similar to what has been shown by \citet{remus:2013}.
More recently, \citet{cappellari:2015b} presented similar results for ETGs from the SLUGGS survey, but their slopes tend to be steeper than those found by \citet{humphrey:2010} and \citet{thomas:2007MNRAS.382..657T}.

\citet{humphrey:2010} also found the central dark matter fraction within the effective radius to vary with other galaxy properties.
They suggest that the accretion of mostly collisionless material leads to a reordering of the system towards an isothermal state \citep[see e.g.,][]{hilz:2012MNRAS.425.3119H}.
Recently, \citet{tortora:2014a} studied the central total mass density profiles of ${\approx}\,4300$ ETGs at $z<0.1$ from the SPIDER survey \citep{labarbera:2010c} and 260 ETGs from the Atlas$^\mathrm{3D}$ survey down to stellar masses as low as $M_*\approx 10^{10}M_\odot$.
They find a clear correlation between the slope of the total central density profile of the galaxies and their size and stellar mass, with their most massive galaxies having slopes close to isothermal ($\gamma \approx -2$).
A correlation of the total density slope within the effective radius and the mass of the ETG has also been reported by \citet{newman:2015}, who studied ETGs in group environments, and by \citet{forbes:2016} from the SLUGGS survey.

As suggested by \citet{remus:2013}, these correlations of the size, dark matter fractions, and in-situ fractions with the total mass density slopes originate from the galaxies' formation histories:
At high redshifts, galaxy growth is dominated by dissipative processes like early gas-rich assembly from filaments and gas-rich mergers.
These processes lead to an enhanced in-situ star formation, which increases the baryonic content of the galaxies especially in their central parts, and thus results in low central dark matter fractions \citep[e.g.,][]{oser:2010ApJ...725.2312O}.
They also lead to steeper total density slopes \citep[see][]{remus:2013,sonnenfeld:2014}.
After a redshift of $z\approx 2$, dissipative processes become less important, while the dissipationless processes start to dominate, i.e., the galaxy growth is mainly driven by merger events and accretion.
This picture is sometimes called the two-phase formation scenario for galaxies \citep[e.g.,][]{naab:2007ApJ...658..710N,guo:2008,naab:2009ApJ...699L.178N,bezanson:2009,oser:2010ApJ...725.2312O,johansson:2012ApJ...754..115J,moster:2013,wellons:2015,furlong:2015,wellons:2016}.
This theoretical framework offers an explanation for the observed trends at $z=0$: 
Gas-poor (collisionless) mergers can increase the dark matter fractions within the effective radius by violent relaxation during major merger events or, even more efficiently, by rapidly increasing the luminous size of the galaxies in minor merger events \citep[e.g.,][]{hilz:2012MNRAS.425.3119H,hilz:2013MNRAS.429.2924H}.
In this picture, ETGs which experienced multiple (mostly) gas poor merger events have lower in-situ fractions, higher dark matter fractions, and larger sizes at $z=0$, but the violent accretion of mostly collisionless material also leads to a reordering of the system towards an isothermal state \citep{johansson:2009,remus:2013}.
Therefore, ETGs that have steeper slopes at $z=0$ had either only a few or only gas-rich accretion events since $z\approx 2$, and therefore still have high in-situ fractions, smaller sizes and low central dark matter fractions.
They also tend to be less massive, however, this tendency is much less pronounced than the correlation between slope and size.
The flattening of the slope with decreasing redshift found in simulations \citep{johansson:2012ApJ...754..115J,remus:2013} can therefore be  naturally explained by ETG growth after about $z\approx 2$ being dominated by mostly dry merger events, which cause a flattening of the density slope while enhancing the central dark matter fraction.

However, observations of galaxies that are strong lenses have revealed a different picture:
Most studies find no evolution of the total density slopes with redshift \citep{koopmans:2006,auger:2010ApJ...724..511A,ruff:2011ApJ...727...96R,bolton:2012ApJ...757...82B,barnabe:2011MNRAS.415.2215B} or even a mild decrease of the density slopes \citep{sonnenfeld:2013} up to redshifts of $z \approx 1$.
\citet{treu:2004}, for example, report average slopes of $\gamma \approx -1.75\pm 0.10$ for their set of 5 ETGs from the LSD survey at redshifts between $0.5<z<1$.

In this paper we use numerical simulations to investigate the evolution of the central mass distributions of ETGs, especially emphasizing the central dark matter fractions.
We include ETGs selected from zoom-in simulations with varying star formation feedback models and ETGs from a large-volume cosmological simulation (Magneticum) including feedback from active galactic nuclei (AGN) in order to understand the impact of those different feedback mechanisms.
Details of the simulations are presented in Sec.~\ref{sec:sims}.
In Sec.~\ref{sec:masssize} we present the evolution of the mass-size relation for the different simulated ETG samples and compare to observations, and in Sec.~\ref{sec:fdm} we discuss the evolution of the central dark matter fractions with redshift and its correlation with galactic mass and size. 
In Sec.~\ref{sec:conspevol} we discuss the total density slopes found in our simulations at different redshifts and their correlations with the dark matter fractions and stellar mass surface densities, and present a comparison with observations.
We conclude in Sec.~\ref{sec:discussevol} with a summary and discussion.

\section{Numerical Simulations}\label{sec:sims}
We use ETGs selected from three different simulation sets to study the effect of different sub-resolution feedback models on our results.
The Magneticum Pathfinder simulations are a set of large-scale hydrodynamical cosmological simulations, while the other two samples are taken from zoom-in re-simulations of selected halo samples.
In the following we briefly describe the simulations with respect to the included feedback models relevant for this work.
For further details, we refer to the previous works that describe the simulations in more detail.
All simulations were performed using extended versions of the parallel TreePM-SPH-code GADGET-2 \citep{springel:2005MNRAS.364.1105S} called P-GADGET-3.
An overview of the general properties of the simulations can be found in Tab.~\ref{tab:sims}.
\def\swd{\hphantom{-}}
\begin{table*}
\begin{center}
\caption{Simulation details}
\label{tab:sims}
\begin{tabular}{l | ccccc}
\hline\hline
                & SPH           & AGN   & Wind  & cooling       & Reference \\
\hline
Oser            & standard      & no    & no    & primordial    & \citet{oser:2010ApJ...725.2312O,oser:2012ApJ...744...63O} \\
Magneticum      & improved      & yes   & weak  & incl. metals  & \citet{hirschmann:2014,teklu:2015paper} \\
Wind            & standard      & no    & strong& incl. metals  & \citet{hirschmann:2013,hirschmann:2015} \\

\hline
\end{tabular}
\end{center}
\end{table*}

\subsection{The Magneticum Pathfinder Simulations}
The Magneticum Pathfinder\footnote{www.magneticum.org} simulations (Dolag et al., in prep.) are a set of hydrodynamical cosmological boxes with volumes ranging from $(2688~\mathrm{Mpc}/h)^3$ to $(48~\mathrm{Mpc}/h)^3$ and resolutions of $m_\mathrm{Gas} = 2.6\times 10^9 M_\odot/h$ up to $m_\mathrm{Gas} = 7.3\times 10^6 M_\odot/h$ \citep[see also][]{bocquet:2016}.
Throughout all simulations, a WMAP7 \citep{komatsu:2011ApJS..192...18K} $\Lambda$CDM cosmology is adapted with $\sigma_8 =0.809$, $h = 0.704$, $\Omega_\Lambda = 0.728$, $\Omega_\mathrm{M} = 0.272$ and $\Omega_\mathrm{B} = 0.0451$ and an initial slope for the power spectrum of $n_\mathrm{s} = 0.963$.

The version of GADGET-3 used for these simulations includes various updates in the formulation of SPH regarding the treatment of the viscosity \citep{dolag:2005MNRAS.364..753D,beck:2015}, thermal conduction \citep{dolag:2004} and the employed SPH kernels \citep{donnert:2013,beck:2015}.
Kinetic feedback from galactic winds is included following \citet{springel:2003MNRAS.339..289S}, and the metal enrichment and star-formation descriptions follow the pattern of metal production from SN Ia, SN II, and AGB mass losses \citep{tornatore:2004MNRAS.349L..19T,tornatore:2007MNRAS.382.1050T}. Each gas particle can form up to four stellar particles. Regarding the metal enrichment, 12 different types of elements are followed.
Additionally, a self-consistent dependence of the gas cooling on the local metallicity is included following \citet{wiersma:2009MNRAS.393...99W}.
Black hole feedback is implemented according to \citet{springel:2005MNRAS.361..776S,fabjan:2010} and \citet{hirschmann:2014}.
For the identification of substructures, a modified version of SUBFIND \citep{springel:2001} is used which includes the contribution of the stars \citep{dolag:2009}.

The Magneticum Pathfinder Simulations show good agreement with observational results for the pressure profiles of the intra-cluster medium \citep{planckmission:2013,mcdonald:2014b} and the properties of the formed black hole (BH) and AGN population over cosmic times \citep{hirschmann:2014,steinborn:2015}.
They have also been found to produce populations of both late and early-type galaxies which resemble the observed angular momentum properties \citep{teklu:2015paper} and global dynamical properties \citep{remus:2015:309}.

In this work we use ETGs selected from a medium-sized $(48~\mathrm{Mpc}/h)^{3}$ cosmological box, which initially contains a total of $2\times576^{3}$ particles (dark matter and gas) with masses of $m_\mathrm{DM} = 3.6\times10^{7} M_{\odot}/h$ and $m_\mathrm{Gas} = 7.3\times10^{6} M_{\odot}/h$.
The mass of the stellar particles varies due to the fact that each gas particle can spawn 4 stellar particles, thus on average the stellar masses are on the order of $m_\mathrm{*} = 2\times10^{6} M_{\odot}/h$.
The gravitational softening evolves from high redshifts to $z=2$, and from $z=2$ to $z=0$ the softening is fixed to $\epsilon_\mathrm{DM} = \epsilon_\mathrm{Gas} = 1.4~\mathrm{kpc}/h$ for dark matter and gas particles and to $\epsilon_\mathrm{*} = 0.7~\mathrm{kpc}/h$ for stellar particles.

\subsection{Cosmological Zoom Simulations}
Cosmological zoom simulations are re-simulations of halos selected from a parent dark-matter-only simulation with higher resolutions and additional baryonic physics included.
While large cosmological volumes provide a statistically representative sample of galaxies, zoom-in re-simulations are computationally less expensive to run and thus can be used to study the impact of the different feedback models on galaxy properties simultaneously allowing for higher resolution levels.

The re-simulations studied in this work are selected from a parent cosmological box of $(72~\mathrm{Mpc}/h)^3$ with $512^3$ DM particles \citep{moster:2010b,oser:2010ApJ...725.2312O}.
The particles have a mass resolution of $m_\mathrm{DM} = 2 \times 10^8 M_\odot/h$, and a comoving gravitational softening of $\epsilon_\mathrm{DM} = 2.52~\mathrm{kpc}/h$ was adopted. For all simulations, a $\Lambda$CDM cosmology based on WMAP3 \citep{spergel:2007ApJS..170..377S} was adopted, i.e., $h = 0.72$, $\Omega_\Lambda = 0.74$, $\Omega_\mathrm{M} = 0.26$ and $\sigma_8 =0.77$.
The used initial power spectrum has a slope of $n_\mathrm{s} = 0.95$.

Halos with masses in the range $10^{11} M_\odot/h$ to $10^{13} M_\odot/h$ were re-simulated with a spatial resolution twice of the original box, leading to $m_\mathrm{DM} = 2.1 \times 10^7 M_\odot/h$, $m_\mathrm{Gas} = M_\mathrm{*} = 4.2 \times 10^6 M_\odot/h$ with gravitational softenings at $z=0$ of $\epsilon_\mathrm{DM} = 0.89~\mathrm{kpc}/h$, $\epsilon_\mathrm{Gas} = \epsilon_\mathrm{*} = 0.4~\mathrm{kpc}/h$
For these re-simulations, a baryon fraction of $f_\mathrm{bar} = 0.169$, that is $\Omega_\mathrm{DM} = 0.216$ and $\Omega_\mathrm{B} = 0.044$, was adopted.

\subsubsection{Oser Simulation Set}
The first set of zoom simulations used in this work were simulated with the code GADGET-2, including star formation and self-regulated supernova feedback according to \citet{springel:2003MNRAS.339..289S} as well as radiative cooling for a primordial mixture of hydrogen and helium \citep{katz:1996ApJS..105...19K}.
In addition, a uniform but redshift dependent UV radiation background was included following \citet{haardt:1996ApJ...461...20H}.
The smaller halos in this simulation set have also been re-simulated with a higher spatial resolution, with particle resolutions of $M_\mathrm{DM} = 3.6 \times 10^6 M_\odot/h$, $m_\mathrm{gas} = m_\mathrm{*} = 7.4 \times 10^5 M_\odot/h$ and comoving gravitational softenings of $\epsilon_\mathrm{DM} = 0.45~\mathrm{kpc}/h$, $\epsilon_\mathrm{gas} = \epsilon_\mathrm{*} = 0.2~\mathrm{kpc}/h$. 
Additional details on the galaxy properties of these simulation can also be found in \citet{oser:2012ApJ...744...63O,hirschmann:2012,remus:2013,naab:2014}.

\subsubsection{Wind Simulation Set}
Several of the zoom-simulations from this sample have been also re-simulated including metal enrichment and momentum-driven galactic winds. For the metal enrichment, contributions from SNIa, SNII and AGB-stars are included following the four elements C, O, Si and Fe following \citet{oppenheimer:2008}.
The values for the mass loading $\eta$ and the wind velocity $v_\mathrm{wind}$ are, however, not constant (as in \citet{springel:2003MNRAS.339..289S}, and thus as in Magneticum), but instead they scale with the galaxy velocity dispersion, motivated by observations of galactic superwinds of \citet{martin:2005,rupke:2005}. 
These simulations are described in detail by \citet{hirschmann:2013,hirschmann:2015}.
In general, the simulations are successful in predicting fairly realistic SFR histories and baryon conversion efficiencies in low-mass halos.
However they still form too massive galaxies with too high SFR due to missing AGN feedback.
From these simulations, the second set of zoom-simulations used in this work are selected.

\subsection{Galaxy Classification}
\def\swd{\hphantom{-}}
\begin{table*}
\begin{center}
\caption{ETG Samples}
\label{tab:etgs}
\begin{tabular}{l | ccccccc}
\hline\hline
   ETG             & Masscut 	& Masscut 	& $N_\mathrm{ETG}/N_\mathrm{gal}$	& $N_\mathrm{ETG}/N_\mathrm{gal}$ 	& $N_\mathrm{ETG}/N_\mathrm{gal}$ 	& $N_\mathrm{ETG}/N_\mathrm{gal}$ & $f(\epsilon_\mathrm{circ})$ \\
   Sample		& ($z=0$) 	& ($z>0$) 	& ($z=0$) 				& ($z=0.5$) 				    & ($z=1$) 				      & ($z=2$)  &  ($|\epsilon_\mathrm{circ}| \leqslant 0.3$)\\
\hline
 Oser           & $M_\mathrm{*} > 5\times 10^{10}M_\odot$  & progenitors                              & 20/\swd21     & 18/\swd21      &  12/\swd21       & 11/\swd21      & 35\% \\
 Magneticum     & $M_\mathrm{*} > 5\times 10^{10}M_\odot$  & $M_\mathrm{*} > 5\times 10^{10}M_\odot$  & 96/269        & 87/305         &  56/298          & \swd5/212      & 40\% \\
 Wind           & $M_\mathrm{*} > 5\times 10^{10}M_\odot$  & (progenitors)                            & \swd5/\swd13  & (\swd2/\swd13) &  (\swd1/\swd13)  & (\swd1/\swd13) & 35\% \\

\hline
\end{tabular}
\end{center}
\medskip

\end{table*}
We use a classification scheme based on the circularity parameter, similar to \citet{scannapieco:2008}:
At first, we align each galaxy along its principal axis of inertia of the stars within $0.1\,R_\mathrm{vir}$.
The stellar mass within $0.1\,R_\mathrm{vir}$ also serves as the definition of the total stellar mass of our galaxies, $M_*$.
The circularity parameter for each gas and star particle within $0.1\,R_\mathrm{vir}$ is then calculated as
\begin{flalign}\label{eq:circularity}
\epsilon_\mathrm{circ} = j_z/j_\mathrm{circ}
\end{flalign}
where $j_z$ is the specific angular momentum of each particle with respect to the $z$-axis and $j_\mathrm{circ}$ is the specific angular momentum expected if the particle were on a circular orbit.
For each galaxy, we then consider the histograms of the circularity parameter.
If a galaxy is rotationally supported, the majority of the stellar (and the gas) particles has a circularity close to $\epsilon_\mathrm{circ} = 1$ (or $\epsilon_\mathrm{circ} = -1$ in case of a counter-rotating (gas) disk).
On the other hand, if a galaxy is dispersion-dominated, the majority of the particles have a circularity close to $\epsilon_\mathrm{circ} = 0$.
In the following, a galaxy is classified as ETG if more than a given fraction $f(\epsilon_\mathrm{circ})$ of its stellar particles have circularities within $-0.3 \leqslant \epsilon_\mathrm{circ} \leqslant 0.3$ and the fraction of cold gas within the half-mass radius (the radius which contains $50\%$ of the stellar mass $M_*$, see also \citealt{remus:2013}) is smaller than $f_\mathrm{gas} <3\%$.
This criterion is a result of a detailed analysis by \citet{teklu:2015paper}, where it was already applied to select `poster child' samples of both spheroidal and disk galaxies in the Magneticum simulations.

As a stringent criterion to avoid contamination of our selected sample of ETGs with disk-like galaxies, merging and distorted structures as well as red disks, we choose a circularity fraction of $f(\epsilon_\mathrm{circ})> 40\%$ based on Fig.~5 and Fig.~19 of \citet{teklu:2015paper}.
This choice of $f(\epsilon_\mathrm{circ})$ also ensures that the same selection criterion can be applied safely at all studied redshifts.
The Magneticum Simulation studied in this work contains a total of 269 halos with a total stellar mass greater than $M_\mathrm{*} > 5\times 10^{10}M_\odot$, of which 96 were classified as ETGs. 
Using the same selection criterion, we also identify ETGs from the Magneticum simulation at higher redshifts, always using the same lower total mass cut of $M_\mathrm{*} > 5\times 10^{10}M_\odot$.
In this work we especially focus on the samples of ETGs selected at $z=0$ (96~ETGs), $z=0.5$ (87~ETGs), $z=1$ (56~ETGs) and $z=2$ (5~ETGs), however, if necessary, additional redshifts are also available.
The selection limits and the numbers of all galaxies and ETGs at all redshifts are listed in Tab.~\ref{tab:etgs}.
The ETGs at high redshift are most likely the progenitors of some of the ETGs at $z=0$, however, we do not trace the ETGs back in time for this analysis but only select them at each timestep according to the above mentioned total mass and circularity criteria.
These ETGs selected from the Magneticum simulations are in the following called `Magneticum ETGs'.

However, if we apply the same choice of $f(\epsilon_\mathrm{circ})> 40\%$ to the zoom simulations, the second set, where the momentum-driven wind model is included, reduces to only 2 galaxies at $z=0$ and none at higher redshifts.
We thus weaken the circularity criterion to have only $35\%$ of the stellar mass with circularities between $-0.3 \leqslant \epsilon_\mathrm{circ} \leqslant 0.3$ (i.e., $f(\epsilon_\mathrm{circ})> 35\%$).
Following \citet{teklu:2015paper}, this weaker criterion should still lead to the selection of dispersion dominated, spheroidal galaxies, but while disk galaxy contamination is still excluded, merging or distorted systems could theoretically contaminate the selection.
Since both samples of zoom simulations are, however, rather small, we simply check all selected galaxies by eye to avoid contamination\footnote{As the number of galaxies in the Magneticum simulation is so much larger than the number of galaxies in both zoom simulation, we do not need to weaken the criterion but rather decided to use the stronger cut to ensure an uncontaminated selected sample of ETGs without the necessity of a visual inspections.}.
The gas criterion, on the other hand, is not changed.

We apply the $f(\epsilon_\mathrm{circ})> 35\%$ criterion to both zoom simulations to ensure a better consistency between both samples, as both samples are simulated from the same set of initial conditions.
For the first set of zoom-simulations \citep{oser:2010ApJ...725.2312O}, nearly all galaxies at $z=0$ were classified as ETGs (20 out of 21 galaxies), with decreasing numbers of ETGs at higher redshifts (see Tab.~\ref{tab:etgs}). 
The ETGs at higher redshifts are always the progenitors of those at lower redshift.
These ETGs are the same as those studied by \citet{remus:2013}, and will in the following be denoted as `Oser ETGs'.

For the second set of zoom-simulations, the number of selected ETGs at $z=0$ enhances to 5 out of 13 galaxies.
At higher redshifts, however, still only one or two galaxies classify as ETG (see Tab.~\ref{tab:etgs}), and thus we only include the 5 ETGs at $z=0$ from this simulation set in our analysis.
These ETGs will be denoted as `Wind ETGs'.
The reduced amount of ETGs clearly shows the impact of the momentum-driven wind model and the metal enrichment, which is discussed in detail in \citet{hirschmann:2013}:
Albeit the initial conditions and the numerical scheme are the same for both the wind and the Oser zoom simulations, the simulations including the momentum-driven wind and metal models produce significantly more disk-like galaxies, while the simulations without those models tend to mostly form ETGs even at high redshifts.
This is due to the more efficient removal of low-angular-momentum gas in the wind models, which later on can be reincorporated in the galaxies, however, then its angular momentum is larger and thus build up more disk-like structures \citep[see also][]{uebler:2014}.

\section{The Mass-Size Relation}\label{sec:masssize}
The mass of an ETG is observationally closely correlated with its size, as shown, for example, by \citet{shen:2003} for SDSS and by \citet{baldry:2012} for the GAMA survey.
\begin{figure*}
  \begin{center}
  \includegraphics[width=0.8\textwidth]{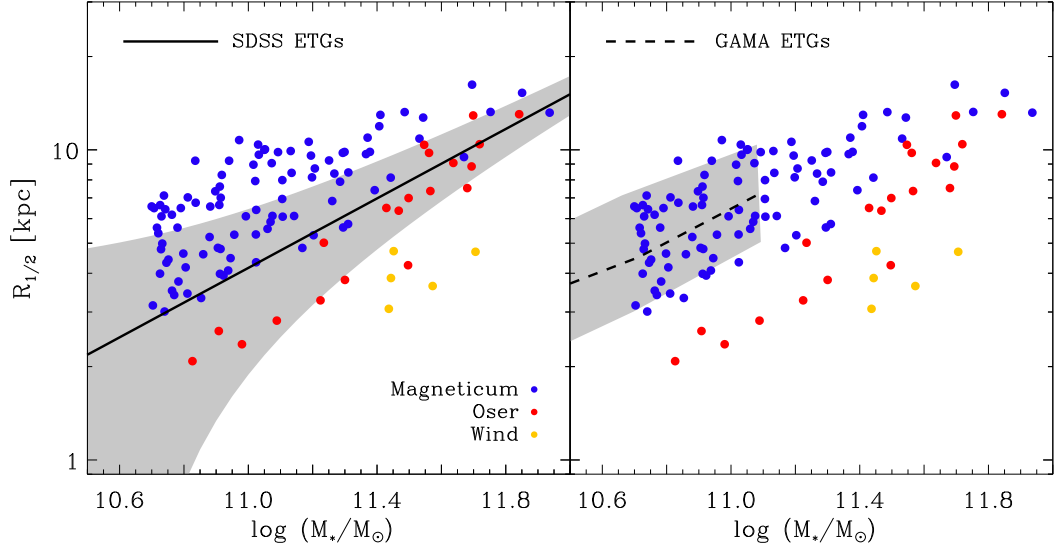}
  \caption{Mass-size relation at $z=0$ for Magneticum ETGs (blue), Oser ETGs (red) and Wind ETGs (yellow) compared to the mass-size relations for ETGs from the SDSS survey \citep[left panel;][black solid line]{shen:2003} and from the GAMA survey \citep[right panel;][black dotted line]{baldry:2012}.
  The shaded areas mark the $1\sigma$-range of observations.
}
  {\label{fig:m_r_z0}}
  \end{center}
\end{figure*}
\begin{figure*}
  \begin{center}
  \includegraphics[width=0.8\textwidth]{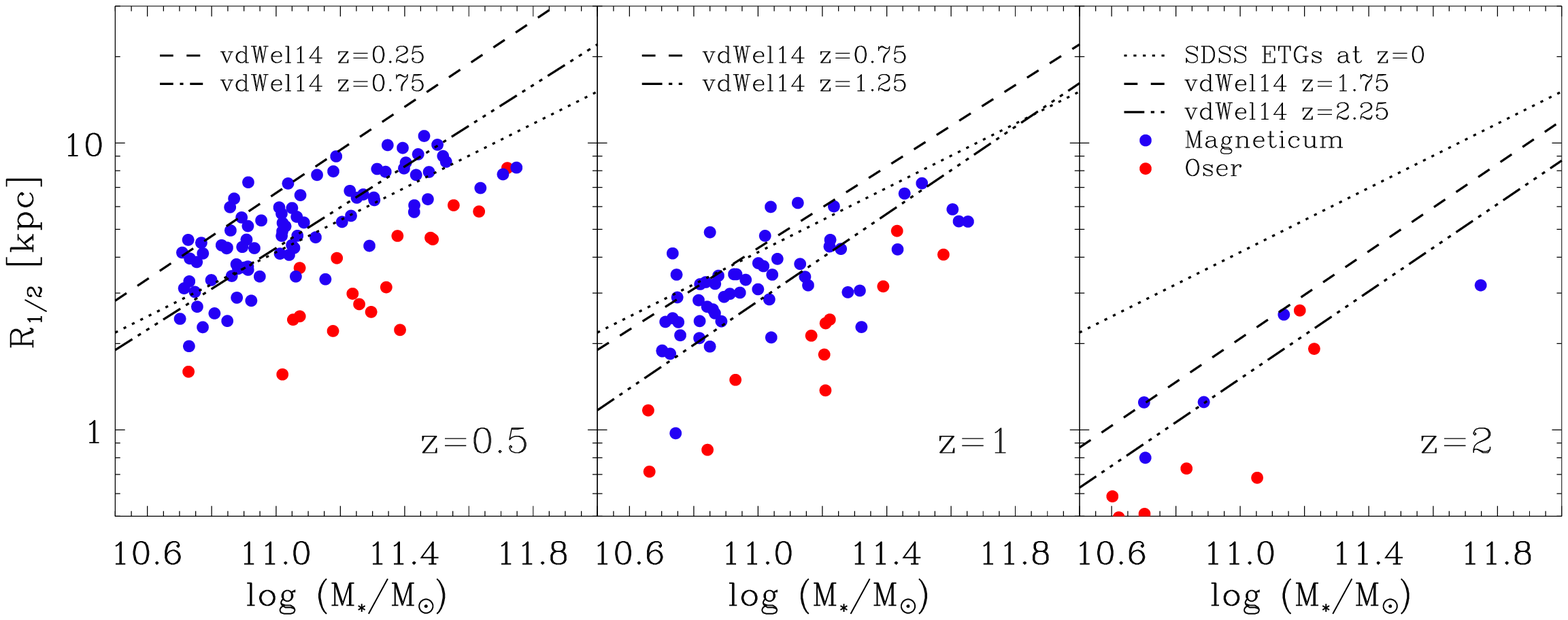}
  \caption{Same as Fig.~\ref{fig:m_r_z0} but at different redshifts (left panel: $z=0.5$; middle panel: $z=1$; right panel: $z=2$).
  The black dotted line shows the relation from SDSS at $z=0$ \citep{shen:2003} for comparison in all panels.
  The black dashed lines in all three panels show the mass-size relations found by \citet{vanderwel:2014} at different redshift bins. In the left panel, we include the relations from \citet{vanderwel:2014} at $z=0.25$ (upper line) and $z=0.75$ (lower line), the central panel includes their relations at $z=0.75$ (upper line) and $z=1.25$ (lower line), and the right panel includes their relations at $z=1.75$ (upper line) and $z=2.25$ (lower line).
}
  {\label{fig:m_r_wz}}
  \end{center}
\end{figure*}
For our simulations, we use the three-dimensional half-mass radius $R_{1/2}$ to compare to the effective radius $r_\mathrm{eff}$ from observations.
We define the three-dimensional half-mass radius $R_{1/2}$ as the radius that contains half of the mass of the galaxy.
To exclude contamination by streams or (disrupted) satellite galaxies, in this calculation we include only stellar particles within 10\% of the virial radius $r_\mathrm{vir}$.
This approach mimics the limitations also experienced by observers as they attempt to observe the low surface brightness component in the outer halo, which is dominated by noise and sky background.
This is the same definition used by \citet{remus:2013} and other similar studies, and it might slightly but systematically overestimate the actual size of the galaxy compared to the observations.

In Fig.~\ref{fig:m_r_z0} we show the mass-size relations from SDSS by \citet{shen:2003} (left panel) and from GAMA by \citet{baldry:2012} (right panel) for their ETGs in comparison with the results from our simulations.
Both the Oser ETGs (red filled circles) and the Magneticum ETGs (blue filled circles) are in good agreement with the SDSS observations at the high mass end, where the GAMA survey does not have any data points.
The Oser ETGs tend to be slightly below the observed relation from SDSS, and are generally more massive at the same sizes than their counterparts from the Magneticum ETG sample.
This most likely originates from the fact that the simulations of the Magneticum ETGs include black holes and their associated feedback, which are not present in the Oser ETG simulations \citep{choi:2015,crain:2015}.

Especially for the smaller masses, the Magneticum ETGs are usually larger in size than the SDSS galaxies of the same mass.
In that mass regime, however, they are in excellent agreement with the observations from the more recent GAMA survey that shows a tilt in the mass-size relation for smaller mass ETGs.
This tilt was also found in a recent re-analysis of the SDSS data by \citet{mosleh:2013}, who showed that the sizes of the smaller galaxies were under-estimated in the previous study \citep[see also][]{hall:2012}.

The ETGs from the Wind ETG sample (yellow filled circles) have masses above $M_\mathrm{*} > 2\times 10^{11}M_\odot$ and are all clearly below the observed mass-size relation.
This is due to the fact that, at the high mass end, the AGN feedback becomes important, and counteracts the effects of the stellar feedback \citep[see also][]{hirschmann:2013}.
The comparison of Magneticum ETGs and Wind ETGs clearly shows that here the AGN feedback is the important missing ingredient \citep[see also][]{dubois:2013}.

At higher redshifts, observations show that ETGs tend to be smaller in size than their present-day counterparts of the same mass \citep[e.g.,][]{vanDokkum:2009Natur.460..717V,bezanson:2013,szomoru:2013,vandesande:2013,vanderwel:2014,marsan:2015}.
Fig.~\ref{fig:m_r_wz} shows the mass-size relation for our simulated ETGs at different redshifts of $z=0.5$ (left panel), $z=1$ (middle panel) and $z=2$ (right panel).
For comparison, we included in each panel the mass-size relations for ETGs from \citet{vanderwel:2014} as dashed black lines and the mass-size relations at $z=0$ from SDSS by \citet{shen:2003} as black dotted line.
All simulations predict the same shift in size towards more compact galaxies at higher redshifts as seen in the observations, albeit we only have very few galaxies at the high mass end at $z\approx 2$ in our sample.
Both, the Oser ETGs and the Magneticum ETGs show the same behaviour, clearly stating that this is independent of the subgrid models for feedback used in the simulations (see Tab.~\ref{tab:sims}).
Nevertheless, both simulations are shifted compared to each other, with the Oser ETGs slightly below the Magneticum ETGs, already at high redshifts.
Again, the Magneticum ETGs are in better agreement with the observations, but at the higher mass end the ETGs from both simulations are more compact than the observed relations from \citet{vanderwel:2014} suggest.
However, note that compact ETGs have also been observed (see for example \citealt{vandesande:2013}).

Another interesting aspect concerning Magneticum ETGs is that the mass growth through dry (minor) merging, which shifts the ETGs strongly in size but not in mass \citep{naab:2009ApJ...699L.178N,hilz:2012MNRAS.425.3119H,hilz:2013MNRAS.429.2924H}, cannot be the only mechanism for the formation of all ETGs that are observed at $z=0$, as the number of Magneticum ETGs is significantly lower at $z=2$ than at $z=0$ and therefore not all those Magneticum ETGs at $z=0$ can be grown from compact ETG progenitors through dry minor merging.
Other processes such as late major merger events or ram-pressure stripping in dense (galaxy cluster) environments can also lead to the formation of ETGs.
Thus, ETGs formed in these ways do not necessarily need to follow the mass-size growth relations predicted for the dry merging scenario.

\begin{figure*}
  \begin{center}
  \includegraphics[width=0.9\textwidth]{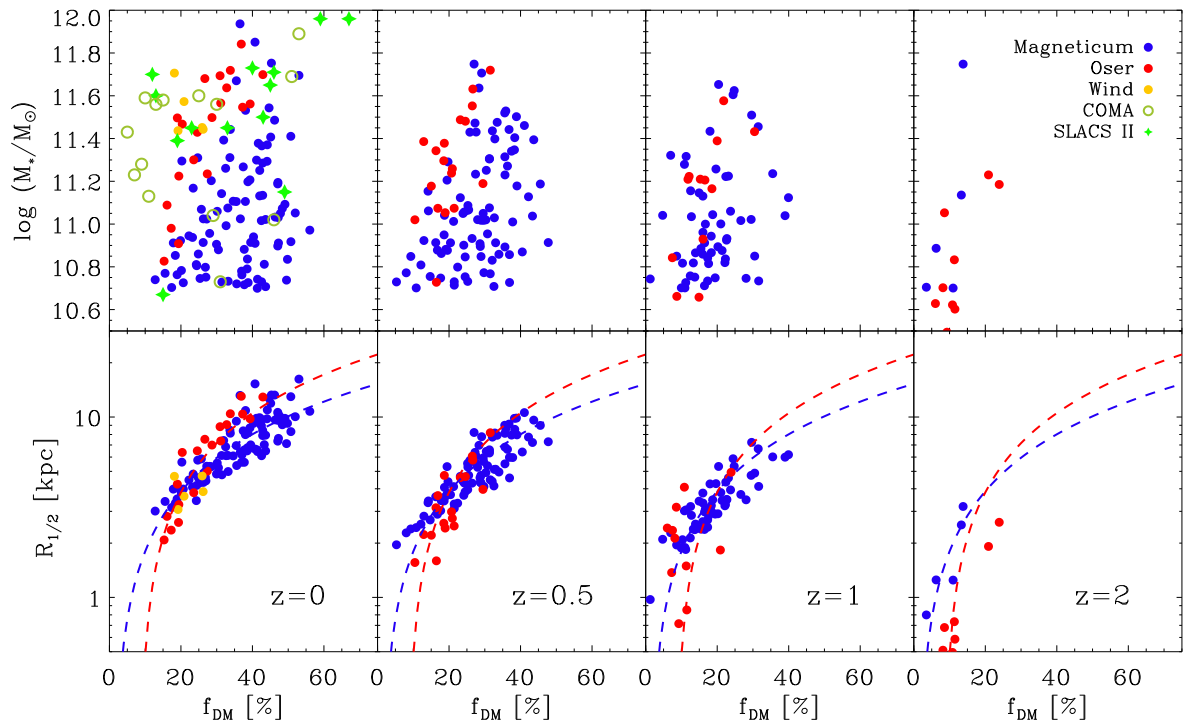}
  \caption{Stellar mass ($M_*$, upper panels) and half-mass radius ($R_\mathrm{1/2}$, lower panels) versus dark matter fractions within the half-mass radius $f_\mathrm{DM}$ for our simulated galaxies.
  Oser ETGs are shown as red filled circles, Wind ETGs as yellow filled circles, and Magneticum ETGs as blue filled circles.
  From left to right: redshift as labeled.
  Upper left panel: For comparison, central dark matter fractions from observations are shown, for Coma cluster ETGs as light green open circles \citep{thomas:2007MNRAS.382..657T}, and for SLACS lenses \citep[][assuming a Chabrier IMF]{barnabe:2011MNRAS.415.2215B} as bright green diamonds.
  Lower panels: Linear fits to the Magneticum ETGs and Oser ETGs are shown in the corresponding colors.
}
  {\label{fig:m_fdm_z}}
  \end{center}
\end{figure*}
\section{Central Dark Matter Fractions}\label{sec:fdm}
We calculate the central dark matter fractions within the stellar half-mass radius, $f_\mathrm{DM}$, for all our galaxies.
At $z=0$ we find a large scatter from less than $f_\mathrm{DM} \approx 10\%$ up to $f_\mathrm{DM} \approx 60\%$.
The Oser ETGs show a much smaller scatter in dark matter fractions than the Magneticum ETGs.
The Wind ETGs populate the same area as the Oser ETGs, indicating that there is no obvious change due to the inclusion of the momentum-driven stellar feedback in this aspect for massive galaxies.
This range is similar to the range of dark matter fractions found by \citet{koopmans:2006} and \citet{barnabe:2011MNRAS.415.2215B} from strong lensing (within different radii), and by \citet{thomas:2007MNRAS.382..657T}, who estimate the central dark matter fractions within the effective radius from dynamical modelling.
It also agrees with the findings by \citet{courteau:2015}, who propose a range of dark matter fractions within the effective radius of $f_\mathrm{DM} \approx 10\%$ up to $f_\mathrm{DM} \approx 60\%$ for ETGs, depending on whether the dark halo has been adiabatically contracted or not.

At $z=0$, we find only a very weak correlation between the stellar mass and the central dark matter fraction for the Magneticum ETGs, but a noticeable tendency for more massive galaxies to have higher central dark matter fractions for the Oser ETGs, as shown in the upper left panel of Fig.~\ref{fig:m_fdm_z}.
The Wind ETGs have slightly lower central dark matter fractions with respect to the stellar mass than both other simulated samples;
however, the sample size is too small to see any possible correlations.

Contrary to the stellar mass, we find a clear correlation between the size of an ETG and its central dark matter fraction at $z=0$:
Smaller ETGs always have smaller central dark matter fractions, as shown in the lower left panel of Fig.~\ref{fig:m_fdm_z}.
This trend is more pronounced for the Oser ETGs than for the Magneticum ETGs.
The Wind ETGs show a similar behaviour as both the Oser ETGs and the Magneticum ETGs.

For comparison, data from \citet{thomas:2007MNRAS.382..657T} are included as light green open circles, and data from \citet{barnabe:2011MNRAS.415.2215B} are shown as green diamonds in the upper left panel of Fig.~\ref{fig:m_fdm_z}.
Similar to the Magneticum ETGs, the observations reveal no clear correlation between the stellar mass and the central dark matter fractions.
There are several massive ETGs in the Coma sample with relatively low dark matter fractions which have no counterpart in any of our simulated samples of ETGs.
This has already been shown for the Oser ETGs by \citet{remus:2013}, and the discrepancy remains with the Magneticum ETGs.
This could be due to a selection effect in both our simulated ETG samples, namely to only include ETGs which are at the centers of dark matter halos and no subhalos, whereas all Coma cluster ETGs in this study are actually substructures within the Coma Cluster host halo.
Those substructures will have suffered from environmental processes like tidal stripping and harassment, which might influence the dark matter fractions even in the central areas \citep{whitmore:1988,dolag:2009,limousin:2009,shu:2015,parry:2016}.
Unfortunately, none of our simulations include a massive galaxy cluster like Coma to test this hypothesis.

At higher redshifts, the dark matter fractions are generally lower than at $z=0$.
At $z=2$, the highest central dark matter fraction for the Magneticum ETGs is $f_\mathrm{DM}\approx 15\%$.
For the Oser ETGs there are two galaxies which have higher fractions, but they are still below $f_\mathrm{DM}<30\%$.
The majority of ETGs in both simulation samples have central dark matter fractions of $f_\mathrm{DM}\approx 10\%$ or less, and the scatter is small.
The average central dark matter fractions at $z=2$ are $\langle f_\mathrm{DM} \rangle = (10\pm5)\%$ and $\langle f_\mathrm{DM} \rangle = (11\pm6)\%$ for the Magneticum and Oser ETGs, respectively, in comparison to $\langle f_\mathrm{DM} \rangle = (36\pm10)\%$ and $\langle f_\mathrm{DM} \rangle = (27\pm8)\%$ at $z=0$.
This agrees with results from \citet{toft:2012} who report dark matter fractions of $f_\mathrm{DM} = (18\pm 20)\%$ for their sample of massive compact quiescent galaxies at $z=2$ and $f_\mathrm{DM} = (46\pm 23)\%$ for their comparison sample of local ETGs.
Similarly, \citet{tortora:2014b} found for their observed ETGs at redshifts up to $z\approx 0.8$ that the high-redshift ETGs have significantly smaller central dark matter fractions than their low redshift counterparts.

The other three upper panels (from left to right) of Fig.~\ref{fig:m_fdm_z} show the stellar mass versus central dark matter fractions for our simulated galaxies at different redshifts, for $z=0.5$, $z=1$, and $z=2$, respectively, and the lower three panels show the halfmass radius versus the central dark matter fraction, correspondingly.
While there is still no clear correlation between stellar mass and central dark matter fractions even at higher redshifts but a general slight tendency for more massive systems to have slightly larger dark matter fractions, the correlation between central dark matter fractions and size is present at all redshifts.
The growth in size and dark matter fractions follows a linear relation $R_\mathrm{1/2} = Af_\mathrm{DM}+ B$ valid for all redshifts, which is shown as blue (red) dashed line for the Magneticum ETGs (Oser ETGs) in all four lower panels of Fig.~\ref{fig:m_fdm_z}.
Those fits were made to the full sample of Magneticum ETGs (Oser ETGs) at all four redshifts together, and we clearly see that, apart from the larger Oser ETGs at $z=2$, all our galaxies are well represented by the lines independent of their redshift
This suggests that the growth mechanisms for the central dark matter fractions and the galaxy sizes are much stronger correlated than with the mass growth..
The fitting parameters $A$ and $B$ for both curves can be found in Tab.~\ref{tab:fits}.
Note that, while the Wind ETGs did not fit at all at the mass-size relation at $z=0$, they nevertheless still fall on the $R_\mathrm{1/2}$--$f_\mathrm{DM}$ relation.

This could be interpreted as a direct effect of the mass growth of galaxies through (minor) merger events:
\citet{hilz:2012MNRAS.425.3119H} showed that an increase in mass by a factor of 2 through minor merger events enhances the central dark matter fraction by about $80\%$, while one equal mass major merger event only increases the dark matter fraction by $25\%$.
Accordingly, the growth in size is much larger in case of multiple minor mergers than for a major merger, thus explaining the much tighter correlation between dark matter fraction and size and the lacking correlation in mass.
Both types of merger can thus also explain the observed increase in central dark matter fractions with decreasing redshift.
In addition, recent merger events can also cause stellar outwards migration \citep{hirschmann:2015}, which would also lead to an increase in the central dark matter fractions.
The fact that this increase is seen in both of our simulation samples again indicates that these trends are not caused by numerical artefacts but are due to the underlying physics of accretion through merger events.

\begin{figure}
  \begin{center}
  \includegraphics[width=\columnwidth]{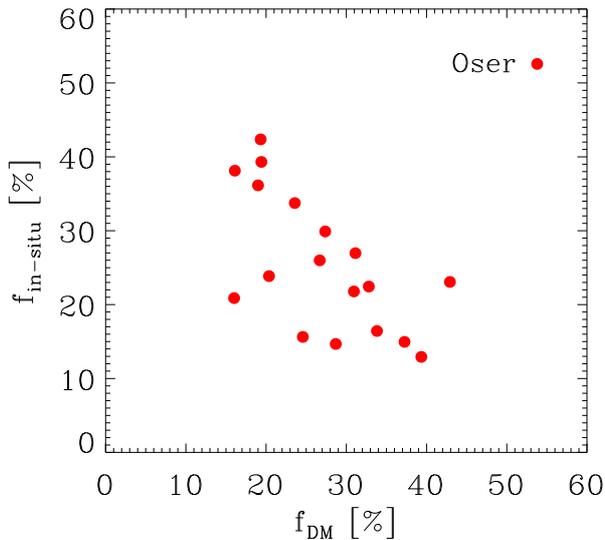}
  \caption{The fraction of stars formed in-situ versus the fraction of dark matter within the half-mass radius for the Oser ETGs at $z=0$.
}
  {\label{fig:fdm_insitu}}
  \end{center}
\end{figure}
This becomes even more clear when we compare the in-situ fraction of the simulated ETGs at $z=0$ with their central dark matter fractions, as shown in Fig.~\ref{fig:fdm_insitu} for the Oser ETGs, using the in-situ fractions obtained by \citet{oser:2012ApJ...744...63O}.
The central dark matter fraction is anti-correlated with the in-situ fraction such that galaxies with lower central dark matter fractions have higher in-situ fractions, while ETGs with high central dark matter fractions have only about $f_\textrm{in-situ}\approx 20\%$ or less.
This again supports the idea that (mostly) dry merging is the main driver of mass growth for those galaxies that were already ETGs at about $z\approx 2$, since dry merging also reduces the fraction of stars formed in-situ while simultaneously enhancing the central dark matter fraction.
But note that the insitu fractions are most likely strongly underestimated in the Oser ETGs, at least compared to abundance matching predictions \citep[see e.g.,][]{hirschmann:2013}.

\section{Total Density Profile Slopes}\label{sec:conspevol}
We fit the total (i.e., stellar plus dark matter) radial density profiles of our ETGs by a single power law with a slope of 
\begin{flalign}
\gamma = \frac{\mathrm{d}\log(\rho)}{\mathrm{d}\log(r)}.
\end{flalign}
These fits have been shown to be a reasonable description of the density profiles for the Oser ETGs by \citet{remus:2013} and for the Magneticum ETGs by \citet{remus:2015:311}.
In the following we will study these slopes in more detail and link them to the quantities studied before, especially the central dark matter fractions.

\subsection{Total Density Slope Evolution with Redshift}
\begin{figure*}
  \begin{center}
  \includegraphics[width=0.7\textwidth]{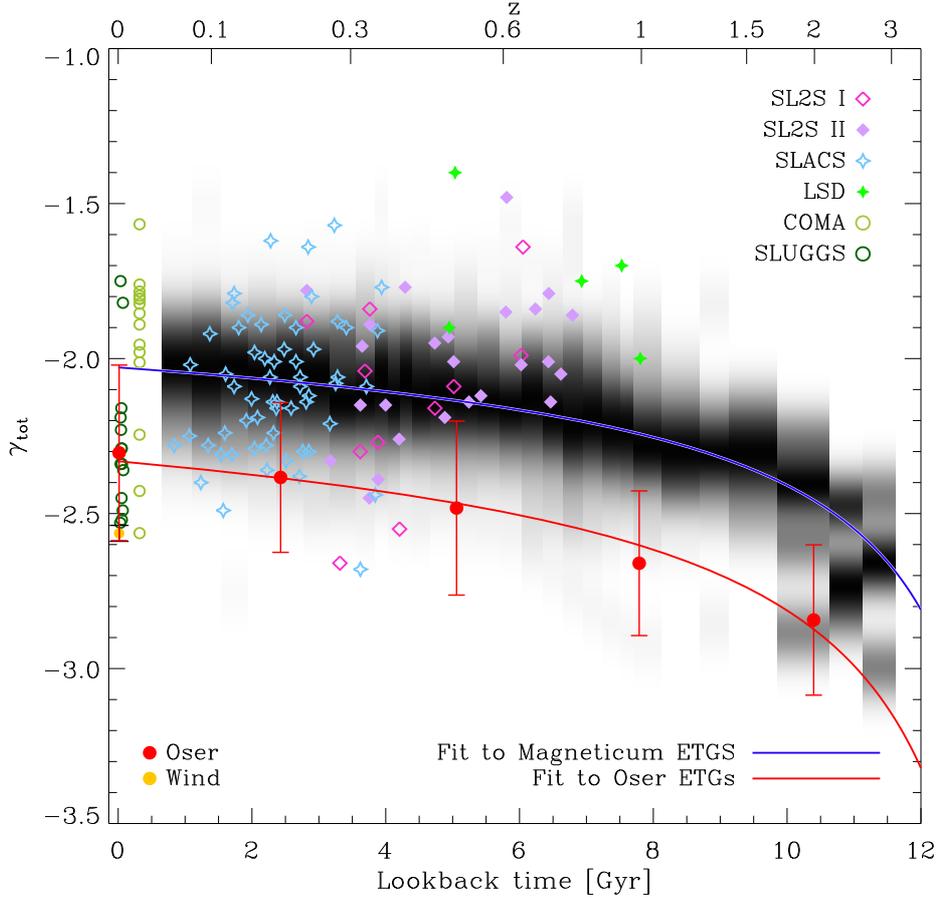}
  \caption{Total density profile power-law slopes versus lookback time for the ETGs from the Magneticum simulation, with total masses above $M_\mathrm{*}\geqslant 5\times 10^{10}M_\odot$ shown as black histograms at each time-bin.
The evolution of the density slope for the Oser ETGs is shown as red filled circles, the average value for the Wind ETGs is shown as yellow filled circle.
For the Magneticum ETGs and Oser ETGs, linear fits in redshift are included in the corresponding color as solid lines (see Tab.~\ref{tab:fits} for the fitting parameters).
The observations are shown as colored symbols:
Light green open circles show the observations for the Coma Cluster ETGs from \citet{thomas:2007MNRAS.382..657T} at $z\approx 0.02$.
Dark green open circles show observations from the SLUGGS survey from \citet{cappellari:2015b} for their set of very nearby galaxies.
All other observations are from strong lensing: SLACS lenses \citep[blue open diamonds,][]{auger:2010ApJ...724..511A}, SL2S lenses (magenta open diamonds: \citealt{ruff:2011ApJ...727...96R}; and lilac filled diamonds: \citealt{sonnenfeld:2013}), and LSD lenses \citep[green filled diamonds,][]{treu:2004}.
}
  {\label{fig:gamma_z_mag}}
  \end{center}
\end{figure*}
At low redshifts we find the total radial density profiles of most of our simulated ETGs to be close to isothermal, i.e., $\gamma_\mathrm{tot} \approx -2$.
This was already shown for the Oser ETGs by \citet{remus:2013}, and turns out to be the case for the Magneticum ETGs as well.
As can be seen in Fig.~\ref{fig:gamma_z_mag}, the Magneticum ETGs tend to have flatter slopes (black histograms), with an average of $\langle\gamma_\mathrm{tot}\rangle = -2.05 \pm 0.13$, than the Oser ETGs, which have a mean value of $\langle\gamma_\mathrm{tot}\rangle = -2.30 \pm 0.28$.
The Wind ETGs have generally steeper density slopes of $\langle\gamma_\mathrm{tot}\rangle = -2.56 \pm 0.03$.
At higher redshifts, both the Oser and Magneticum ETGs have generally steeper slopes than at low redshifts, although the Oser ETGs always show steeper slopes than the Magneticum ETGs.
Nevertheless, even if the actual values for the mean total density slopes for the Oser ETGs are smaller than for the Magneticum ETGs at each redshift bin, the general evolutionary trends are the same.
As shown by the red and blue lines in Fig.~\ref{fig:gamma_z_mag}, the evolution of the average total density slopes $\langle\gamma_\mathrm{tot}\rangle$ in both simulated samples of ETGS follows a linear correlation with $z$, i.e., $\langle\gamma_\mathrm{tot}\rangle = A\,z + B$, with very similar values for $A$ found for both simulations ($A = -0.21$ for the Magneticum ETGs, $A = -0.27$ for the Oser ETGs, see also Tab.~\ref{tab:fits} for the fit parameters).
A similar evolutionary trend was also found by \citet{johansson:2012ApJ...754..115J} for their set of re-simulations.
The generally flatter average slopes of the Magneticum ETGs compared to the Oser ETGs at all redshifts originate most likely from the additional feedback included in the Magneticum simulations, especially the AGN feedback (see also \citealt{dubois:2013}).
However, the different feedback models do not cause a change in the general trends of the total density slopes with redshift.
\def\swd{\hphantom{-}}
\begin{table}
\begin{center}
\caption{Fitting parameters}
\label{tab:fits}
\begin{tabular}{l|ccc}
\hline\hline
    & & Oser & Magneticum \\
    & & ETGs & ETGs \\
\hline
\multirow{2}{*}{$\gamma_\mathrm{tot} = Az + B$}		 		& A & $-0.27$ 	& $-0.21$ \\
									& B & $-2.33$ 	& $-2.03$\\ 
\hline
\multirow{2}{*}{$R_\mathrm{1/2} = Af_\mathrm{DM}+ B$} 			& A & $\swd 0.33$  	& $\swd 0.21$ \\
									& B & $ -2.87$  	& $-0.28$\\
\hline
\multirow{2}{*}{$\gamma_\mathrm{tot} = A\,f_\mathrm{DM} + B$} 		& A & $\swd 0.03$      & $\swd 0.01$\\
									& B & $-3.20$          & $-2.52$ \\
\hline
\multirow{2}{*}{$\gamma_\mathrm{tot} = A\,\log(\Sigma_*) + B$}          & A & $-0.57$          & $-0.38$\\
									& B & $\swd 3.06$      & $\swd 1.32$ \\
\hline
\multirow{2}{*}{$f_\mathrm{DM} = A\,\log(\Sigma_*) + B$}          	& A & $-0.05$ 		& $-0.03$ \\
									& B & $10.77$ 		& $\swd9.73$ \\
\hline
\multirow{2}{*}{$\gamma_\mathrm{tot} = A\,\log(R_\mathrm{1/2}) + B$}	& A & $\swd0.71$	& $\swd0.69$ \\
									& B & $-2.83$ 		& $-2.61$ \\
\hline
\end{tabular}
\end{center}
\medskip
Dark matter fractions $f_\mathrm{DM}$ are calculated within the half-mass radius $R_{1/2}$. The parameters $A$ and $B$ are in the according units of $M_\odot$ and kpc.\\
\end{table}

While both our simulated samples of Magneticum ETGs and Oser ETGs clearly show that at higher redshifts the total density slopes $\gamma_\mathrm{tot}$ were steeper than at lower redshifts, the observations indicate a different behaviour.
\citet{sonnenfeld:2013} demonstrated that the power-law slopes of the radial density profiles inferred from observations of strong lensing ETGs are flatter at higher redshifts than at low redshifts.
In a subsequent paper, \citet{sonnenfeld:2014} argued that merger events at low redshifts thus must contain a significant amount of cold gas to steepen the slope.
Fig.~\ref{fig:gamma_z_mag} highlights the tension between simulations and observations, showing the observations as colored symbols.
While the total density slopes found for the Oser ETGs at high redshifts are all much steeper than the observed ones, the Magneticum ETG sample actually includes galaxies with total density slopes as flat as the observed ones even at higher redshifts.
This is mainly due to the fact that the Magneticum ETG sample is selected from a full cosmological box, and thus also includes massive evolved galaxies even at redshifts as high as $z=2$ which are the progenitors of the most massive galaxies at $z=0$.

Even though there are ETGs in the Magneticum ETG sample which show similarly flat slopes as the observations at high redshifts, they are nevertheless outliers.
The overall trends between simulations and observations differs strongly at high redshifts.

\subsection{Mock Observations of the \mbox{Total Density Slopes}}
Since the simulations and observations are in good agreement regarding general properties of the spheroidals such as masses, dark matter fractions, and sizes at different redshifts, and only diverge in the interpretation of the evolutionary trends for the total density slopes, the question remains whether this could be explained by a methodical issue.
To answer this question, we try to reproduce the observational method of determining density slopes by using the simulated spheroidals as ``fake'' lens galaxies and analyzing these galaxies using the same tools as the observers.
A.~Sonnenfeld kindly provided us with the analysis program used to calculate the total density slopes from the SL2S observations.
The input information required by this method is the effective radius $R_\mathrm{eff}$ of the lens galaxy, the Einstein radius of the lens $R_\mathrm{Ein}$, the projected line-of-sight velocity dispersion $\sigma_\mathrm{LOS}$ within $0.5\,R_\mathrm{eff}$, and the total mass $M_\mathrm{tot}$ within the lens area.

To make mock observations from our simulations, we use the following inputs to meet those requirements:
\begin{enumerate}
\item The effective radius $R_\mathrm{eff}$ of the lens galaxy is (as discussed before) approximated by the stellar half-mass radius $R_{1/2}$.
\item For the Einstein radius of the lens we assume $R_\mathrm{Ein}=1.5\,r_{1/2}$,
according to the ratios of $R_\mathrm{Ein}$ to $R_\mathrm{eff}$ which have been found by \citet{sonnenfeld:2013b} and \citet{ruff:2011ApJ...727...96R} for the SL2S survey.
The ratios of the lenses studied in the SLACS (and BELLS) survey were usually smaller, but the ratios for the LSD survey were of similar order.
This choice, however, should not strongly influence the results, since \citet{sonnenfeld:2013} showed that the ratio between effective radius and Einstein radius does not change the resulting slopes significantly
(the influence of changes in this ratio on the resulting total density slopes was smaller than the error of the measurements).
This is important since there is a tendency for lenses at higher redshifts to have larger ratios of $R_\mathrm{Ein}$ to $R_\mathrm{eff}$ due to geometrical reasons.
\item To obtain the projected line-of-sight velocity dispersion $\sigma_\mathrm{LOS}$ within $0.5\,R_\mathrm{eff}$,
we rotate our spheroidals to both a face-on and an edge-on projection,
and calculate the line-of-sight velocity dispersion within half of the half-mass radius for both projections separately.
In the following study we will always consider both projections, as they are the maximum and minimum values that can be found.
\item For the total mass within the lens area we include all stellar, gas, and dark matter particles within the given projected radius of $R_\mathrm{Ein}=1.5\,R_{1/2}$, for both projections.
\end{enumerate}

\begin{figure}
  \begin{center}
  \includegraphics[width=\columnwidth]{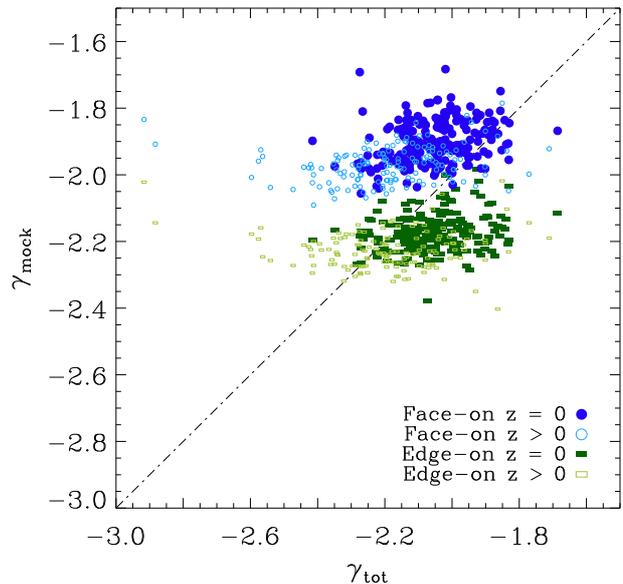}
  \caption{Total mass density profile slopes calculated for the mock lensing observations of the simulated Magneticum ETGs, $\gamma_\mathrm{mock}$,
against the intrinsic total mass density profile slopes, $\gamma_\mathrm{tot}$.
The dash-dotted line shows the 1:1 relation.
The blue circles show the values for the face-on projections, the green bars those for the edge-on projections.
Open symbols represent the values at higher redshifts, filled symbols those at $z=0$.
}
  {\label{fig:mock_lensing}}
  \end{center}
\end{figure}

\begin{figure}
  \begin{center}
  \includegraphics[width=\columnwidth]{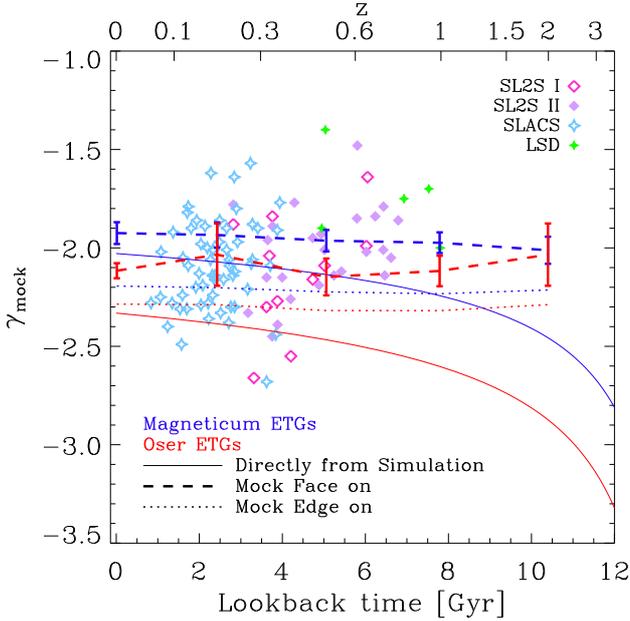}
  \caption{Same as Fig.~\ref{fig:gamma_z_mag} but for the mock observations of our simulations.
The dashed lines show the average radial density slopes obtained for face-on projections,
the dotted lines show the same for the edge-on projections of the Magneticum ETGs (blue) and the Oser ETGs (red).
For comparison, the solid lines again show the linear fits in redshift to the intrinsic density slopes from the simulations,
and the colored symbols are the same strong-lensing observations as in Fig.~\ref{fig:gamma_z_mag}.
}
  {\label{fig:mock_lensing_z}}
  \end{center}
\end{figure}

\begin{figure*}
  \begin{center}
  \includegraphics[width=\textwidth]{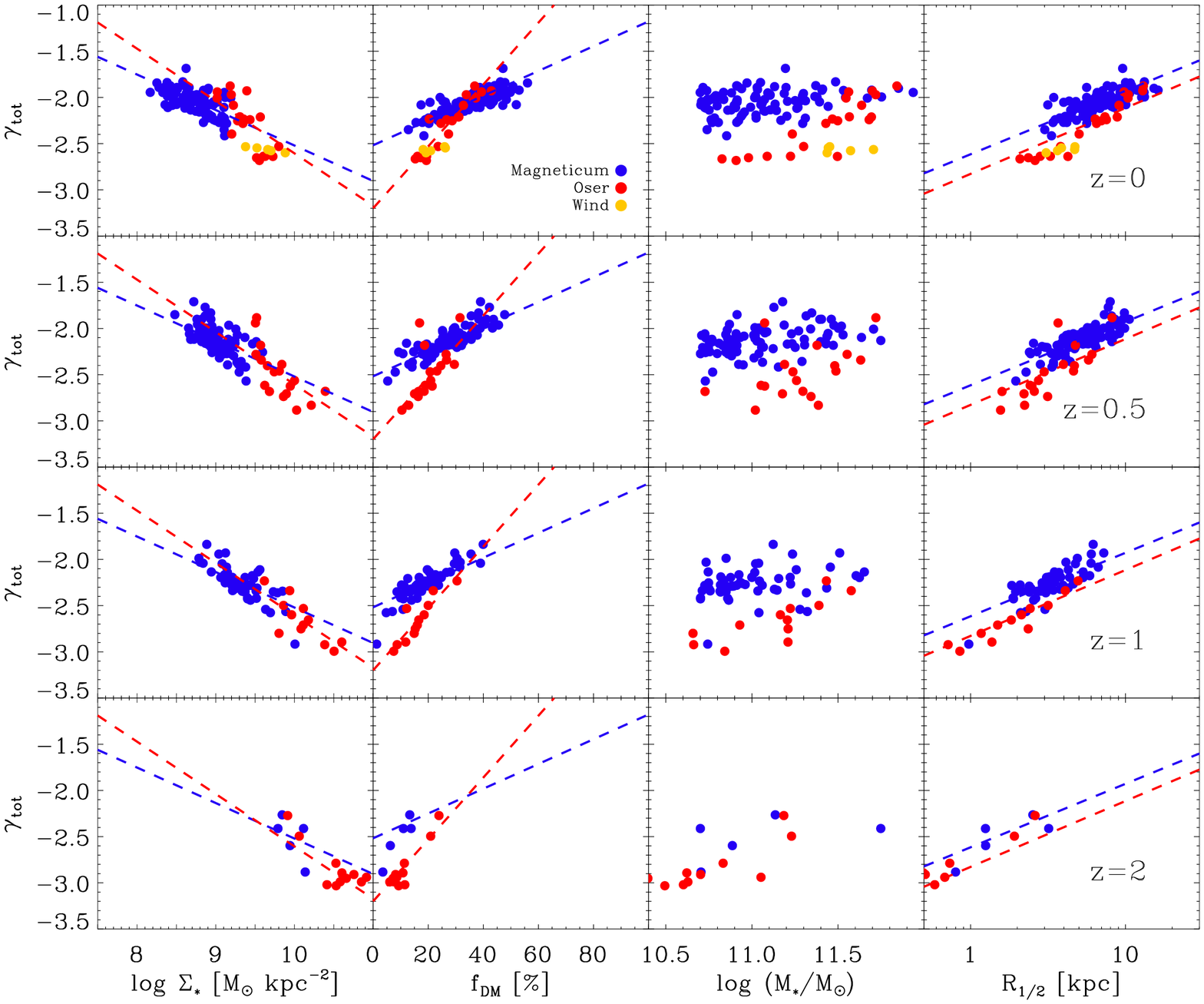}
  \caption{Total mass density profile slopes $\gamma_\mathrm{tot}$ versus stellar mass surface densities $\Sigma_*$ (first column), versus central dark matter fractions $f_\mathrm{DM}$ (second column), versus stellar mass $M_*$ (third column), and versus half-mass radius $R_\mathrm{1/2}$ (fourth column).
  Rows: From top to bottom: $z=0$, $z=0.5$, $z=1$, $z=2$.
  Magneticum ETGs are shown as blue circles, Oser ETGs as red circles, and Wind ETGs as yellow circles.
  Dashed blue and red lines show fits to the Magneticum ETGs and Oser ETGs, respectively, with the fit parameters listed in Tab.~\ref{tab:fits}.
}
  {\label{fig:gamma_sigma_fdm_z}}
  \end{center}
\end{figure*}

The density slopes we obtain from these mock observations are shown in Fig.~\ref{fig:mock_lensing},
for the edge-on (green bars) and for the face-on (blue circles) view.
Galaxies at $z=0$ are shown as filled symbols, those at higher redshifts ($z=0.5$, $z=1$, and $z=2$) as open symbols.
We only show the result for the Magneticum ETGs as the behaviour is the same for the Oser ETGs.

As can clearly be seen, there is a strong discrepancy between the resulting density slopes taken directly from the simulations,
$\gamma_\mathrm{tot}$, and the ``observed'' density slopes, $\gamma_\mathrm{mock}$.
Ideally, the data points should lie along the dash-dotted line, i.e., $\gamma_\mathrm{mock} \approx \gamma_\mathrm{tot}$.
However, the mocked slopes are closely scattered around $\gamma_\mathrm{mock} =-2.2$ for the edge-on and $\gamma_\mathrm{mock} = -1.9$ for the face-on projections,
while the intrinsic slopes from the simulations show a much larger range of variation from $\gamma_\mathrm{tot}\approx -3$ to $\gamma_\mathrm{tot}\approx -1.6$.
The discrepancy between the ``real'' total density slopes and the mocked slopes is most prominent at the steeper slope end.
This behaviour might be related to the results presented by \citet{vandeven:2009}, who model the lensing properties of galaxies with different density distributions and find that non-isothermal density profiles may appear isothermal if measured at the Einstein radius.

Indications for such a discrepancy between simulated and observed slopes had already been presented by \citet{sonnenfeld:2014} (especially their Fig.~8)
using a comparison sample of major-merger simulations.
However, the difference is much larger for our cosmological simulations where we find significantly steeper slopes at high redshifts.
In addition, as shown for the major-merger sample studied by \citet{remus:2013}, with the usual present-day configuration for disk galaxies, the initial slopes are also close to isothermal, which is why the effect of the discrepancy between the mocked and intrinsic slopes is less pronounced in those cases.
Given that the line-of-sight velocity dispersion is one of the major input parameters in calculating the mocked slopes, and actually the most error-prone one,
we tested whether the difference between the mocked and the intrinsic slopes depends on the line-of-sight velocity dispersions of our simulated sample,
but we did not find any correlation.

To conclude this analysis, in Fig.~\ref{fig:mock_lensing_z} we again plot the evolution of the slopes as in Fig.~\ref{fig:gamma_z_mag}, but this time including the mock observations of our simulations.
The dashed lines show the median values obtained for the face-on projections, the dotted lines those for the edge-on views at the different redshifts.
As before, the solid lines show the intrinsic evolutionary trends for the Magneticum ETGs (blue) and Oser ETGs (red) as measured directly from the simulations.
At $z=0$, the intrinsic slopes from the simulations are close to isothermal and thus there is little difference between the mocked slopes $\gamma_\mathrm{mock}$ and the intrinsic slopes $\gamma_\mathrm{tot}$.
At higher redshifts, however, where the intrinsic slopes are much steeper, a discrepancy between the mocked and intrinsic average values becomes apparent, in accord with the behaviour seen in Fig.~\ref{fig:mock_lensing}.
Thus, the slopes found for the mock observations show no evolution trend with redshift in contrast to the intrinsic slopes directly derived from the simulations.

Most importantly, both simulation sets now show not only the same values as the observations at low redshifts, but also the same behaviour with redshift, namely that there is basically no evolution of the density slopes.
While this identifies the origin of the discrepancies between the evolutionary trends from observations and simulations,
it raises the question of how the analysis tools used to infer the total density slopes from strong-lensing measurements can be improved.
A more detailed comparison of the observational methods with simulations will hopefully help to solve those issues,
and enhance our understanding of the interaction processes between the dark and luminous components as well as the role of the cold gas in establishing the total radial density profiles.

\subsection{Correlating Galaxy Properties with the \mbox{Total Density Slope}}
The central dark matter fractions $f_\mathrm{DM}$ correlate with the in-situ fractions $f_\textrm{in-situ}$ of the ETGs (Fig.~\ref{fig:fdm_insitu}), and the in-situ fractions show a correlation with the total density slope $\gamma_\mathrm{tot}$ (Fig.~10 in \citealt{remus:2013}).
Therefore, there should also exist a correlation between the central dark matter fractions and total density slope $\gamma_\mathrm{tot}$.
The second column of Fig.~\ref{fig:gamma_sigma_fdm_z} shows $\gamma_\mathrm{tot}$ versus $f_\mathrm{DM}$, for the Magneticum ETGs (blue circles) and the Oser ETGs (red circles) at different redshifts ($z=0$, $z=0.5$, $z=1$, and $z=2$ from top to bottom).
We find a clear correlation between both quantities for both samples of ETGs, in the sense that ETGs with a flatter slope have larger central dark matter fractions.
However, the correlations have very different slopes:
While the Oser ETGs show a steep, nearly linear increase in $\gamma_\mathrm{tot}$ with increasing $f_\mathrm{DM}$, the Magneticum ETGs show a strong increase in $\gamma_\mathrm{tot}$ for small changes in $f_\mathrm{DM}$ at low central dark matter fractions, and a flattening of the correlation above $f_\mathrm{DM}\approx 20\%$, where the slopes only change slightly, and are on average already close to isothermal.

\begin{figure*}
  \begin{center}
  \includegraphics[width=\textwidth]{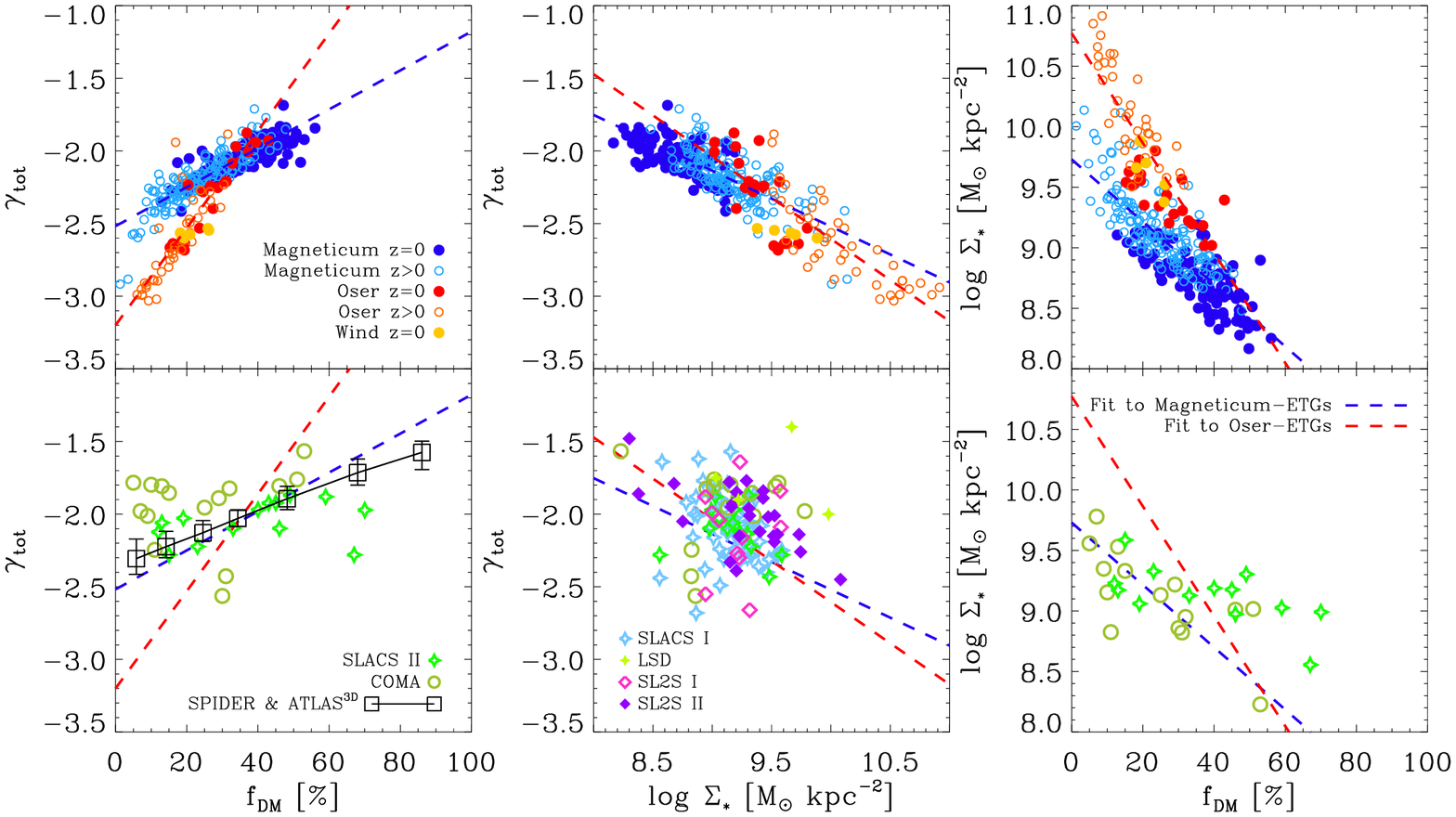}
  \caption{Correlations between central dark matter fractions $f_\mathrm{DM}$, total mass density slopes $\gamma_\mathrm{tot}$ and stellar mass surface density $\Sigma_*$.
  Left panels: Total density slopes $\gamma_\mathrm{tot}$ versus central dark matter fractions $f_\mathrm{DM}$.
  Middle panels: Total density slopes $\gamma_\mathrm{tot}$ versus stellar mass surface density $\Sigma_*$.
  Right panels: Stellar mass surface density $\Sigma_*$ versus central dark matter fractions $f_\mathrm{DM}$.
  The upper panels show the simulations: Magneticum ETGs at $z=0$ (blue filled circles) and $z>0$ (light blue open circles), Oser ETGs at $z=0$ (red filled circles) and $z>0$ (open light red circles), and Wind ETGs at $z=0$ (yellow circles).
The blue and red dashed lines represent fits to the total populations of the Magneticum ETGs and Oser ETGs at all four redshifts, respectively; see Tab.~\ref{tab:fits} for the parameters of the fits.
  The lower panels show observations on top of the fits to the Magneticum ETGs and Oser ETGs from the upper panels.
Black squares are taken from the SPIDER and ATLAS$^\mathrm{3D}$ surveys from \citet{tortora:2014a}, light green open circles show the results for the COMA cluster ETGs from \citep{thomas:2007MNRAS.382..657T}, and green open diamonds mark the results for SLACS lenses \citep{barnabe:2011MNRAS.415.2215B}.
In the central panel, additional observational results from strong lensing are included: SLACS lenses from \citet{auger:2010ApJ...724..511A} (light blue diamonds), SL2S lenses from \citet{ruff:2011ApJ...727...96R} (magenta open diamonds) and \citet{sonnenfeld:2013} (purple filled diamonds), and LSD lenses from \citet{treu:2004} (filled light green diamonds).
}
  {\label{fig:gamma_sigma_fdm}}
  \end{center}
\end{figure*}

Another interesting observational quantity to compare to is the effective stellar mass surface density, defined as
\begin{flalign}\label{eq:massdens}
\Sigma_* = \frac{M_*}{2\pi r_\mathrm{eff}^2}
\end{flalign}
following \citet{sonnenfeld:2013}, which can be considered as a measurement of the concentration of the stellar component:
the smaller $\Sigma_*$, the less concentrated a galaxy.
We calculated $\Sigma_*$ for our simulated halos, using the stellar half-mass radius instead of the effective radius in Eq.~\ref{eq:massdens}, and find a clear correlation with the total density slope (see left column of Fig.~\ref{fig:gamma_sigma_fdm_z}) for both simulation samples:
More compact galaxies have steeper total density slopes.
Most of the Oser ETGs are much more concentrated than the Magneticum ETGs.
This is due to the missing AGN feedback in the Oser ETG simulations, as the AGN feedback suppresses the star formation in the center of the galaxies.
A similar result has been reported by \citet{dutton:2015} who use artifical quenching instead of AGN feedback to suppress the star formation.
The Wind ETGs, which also do not have AGN feedback, are similarly compact as the Oser ETGs;
however, their total density slopes are all very steep, which shows that the stellar feedback has a much stronger influence on the total density slope than on the concentration.

The third column of Fig.~\ref{fig:gamma_sigma_fdm_z} shows the total density slope $\gamma_\mathrm{tot}$ versus the stellar mass $M_*$ at $z=0$, $z=0.5$, $z=1$, and $z=2$ (from top to bottom).
We find a much weaker trend between the slopes of the total density profiles and the stellar masses for the Magneticum ETGs than for the Oser ETGs.
This holds true for all redshifts, and is also a result of the additional AGN feedback that is included in the Magneticum ETGs.
In this correlation, the influence of the stellar feedback can be seen best, as the Wind ETGs here clearly deviate from the correlations found for the Oser ETGs, namely their total density profiles are steeper for a given stellar mass, while they agree with the Oser ETGs for all the other correlations at $z=0$.
This is due to the fact that the stellar feedback enhances the fraction of stars formed in situ while simultaneously lowering the fraction of accreted stars \citep[see][]{hirschmann:2013}, and higher in situ fractions correlate with steeper total density slopes \citep[see][]{remus:2013}.
On the contrary, the AGN feedback has the opposite effect of the stellar feedback: 
While it suppresses the star formation at the center, it enhances the amount of accreted stars at lower redshifts \citep[e.g.,][]{dubois:2013}, and also leads to flatter total density slopes at a given stellar mass.

For the half-mass radii $R_\mathrm{1/2}$ we see a clear correlation with the total density slopes $\gamma_\mathrm{tot}$ for both simulations.
As shown by the rightmost column of Fig.~\ref{fig:gamma_sigma_fdm_z}, at all redshifts (from $z=2$ to $z=0$, bottom to top) the galaxies are distributed along the correlation defined by the overall evolution (colored lines) which can be described as $\gamma_\mathrm{tot} = A\,\log(R_\mathrm{1/2}) + B$.
Interestingly, both the Magneticum and Oser ETGs lead to almost identical correlations, differing only by a small offset $B$ (see Tab.~\ref{tab:fits}).
The Wind ETGs at $z=0$ are again roughly in agreement with the Oser ETGs, indicating that the offset in the relations found for the Magneticum ETGs and the Oser ETGs originates from the included AGN and not from the stellar feedback.

A similar behaviour can be seen for the evolution of the correlation between the total density slopes $\gamma_\mathrm{tot}$ and the stellar mass surface density $\Sigma_*$ as well as the correlation between the total density slopes $\gamma_\mathrm{tot}$ and the central dark matter fractions $f_\mathrm{DM}$.
Both the Magneticum ETGs and Oser ETGs evolve with redshift along a path that can be estimated by fitting to all galaxies at all redshifts for the respective simulation sets.
The correlations can be well described by $\gamma_\mathrm{tot} = A\,\log(\Sigma_*) + B$ and $\gamma_\mathrm{tot} = A\,f_\mathrm{DM} + B$ (fitting parameters see Tab.~\ref{tab:fits}, and blue and red dashed lines in the first and second rows of Fig.~\ref{fig:gamma_sigma_fdm_z}).
This can also be seen in the upper rows of Fig.~\ref{fig:gamma_sigma_fdm}, where we show $\gamma_\mathrm{tot}$ versus $f_\mathrm{DM}$ (left column) and $\gamma_\mathrm{tot}$ versus $\Sigma_*$ (middle column), including the Magneticum ETGs and Oser ETGs at all four redshift bins.
The fits are a good representation of the galaxies at all redshifts, and we again see the co-evolution of all three quantities with redshift.
The Wind ETGs show a similar behaviour to the Magneticum ETGs and Oser ETGs, however, they tend to follow the fits found for the Oser ETGs and not those for the Magneticum ETGs.
This suggests that the slight differences seen for the trends for Magneticum ETGs and Oser ETGs originate mostly from the impact of the included AGN feedback and not from the stellar feedback.
For completeness, we show the stellar mass density $\Sigma_*$ versus the central dark matter fraction $f_\mathrm{DM}$ in the upper right panel of Fig.~\ref{fig:gamma_sigma_fdm}.
As expected, there is a clear tendency for ETGs with larger $f_\mathrm{DM}$ to be less compact, and while this is supported by both the Magneticum ETGs and the Oser ETGs, there is a clear difference between the slopes of the two simulations.
Of these three evolutionary trends, the correlation between $f_\mathrm{DM}$ and $\gamma_\mathrm{tot}$ is the tightest, and is thus best suited for comparison to observations to constrain different AGN feedback models.

These evolutionary trends support the idea that, after about $z=2$, the evolution of ETGs becomes more dominated by merger events, which enhance the central dark matter fractions, lead to stronger growth in size than in mass, and evolve the total density slope towards an isothermal solution through dynamical friction and violent relaxation.
In summary, we find that simulations show clear evidence for a co-evolution of $\gamma_\mathrm{tot}$, $\Sigma_*$, and $f_\mathrm{DM}$ with a relation that is independent of redshift.
While the relations are similar, the slopes for these correlations are different for different sets of simulation models, and therefore we suggest that they can be used as a test for different feedback models.

\subsection{Comparison to Observations}
We compare the correlations found between $\gamma_\mathrm{tot}$, $f_\mathrm{DM}$, and $\Sigma_*$ to observations from strong lensing as well as dynamical modelling.
This is shown in the lower panels of Fig.~\ref{fig:gamma_sigma_fdm}, together with the correlations for the Magneticum ETGs and Oser ETGs (blue and red dashed lines, respectively, as in the upper three panels of the same figure).

The lower left panel of Fig.~\ref{fig:gamma_sigma_fdm} shows $\gamma_\mathrm{tot}$ versus $f_\mathrm{DM}$.
The observations from strong lensing \citep[SLACS sample,][]{barnabe:2011MNRAS.415.2215B} and  dynamical modelling \citep[COMA cluster ETGs,][and SPIDER and ATLAS$^\mathrm{3D}$, \citealt{tortora:2014a}]{thomas:2007MNRAS.382..657T} are included as open green diamonds, open light green circles and open black squares, respectively.
They reveal a very good agreement with the correlation found for the Magneticum ETGs, while their match with the Oser ETGs is rather poor.
The strongest deviations are present for the Coma Cluster ETGs, which show a much larger scatter than the other observations and our simulations, as they include a subset of ETGs which have extremely high masses and flat total density slopes but low central dark matter fractions.
As discussed before, this could be an environmental effect, since those galaxies all reside inside a massive galaxy cluster environment as satellite galaxies.
The Magneticum ETGs are in excellent agreement with the observations from SPIDER and ATLAS$^\mathrm{3D}$ \citep{tortora:2014a}.
This result clearly highlights the importance of AGN feedback for the evolution of ETGs.

The middle panel of Fig.~\ref{fig:gamma_sigma_fdm} shows $\gamma_\mathrm{tot}$ versus $\Sigma_*$ compared to all strong lensing observations (see Fig.~\ref{fig:gamma_z_mag}) in addition to the Coma Cluster ETGs from \citet{thomas:2007MNRAS.382..657T}.
Generally, \citet{sonnenfeld:2013} but also \citet{auger:2010ApJ...724..511A} report for their observations, that ETGs with more concentrated stellar components have steeper total density profiles well in agreement with the correlation found in the simulations.
The trend seen for the observations is much weaker than the relations found from both of our simulations, but it agrees best with the results from strong lensing presented by \citet{sonnenfeld:2013} (purple filled diamonds), who find the same slope $A$ for their fit as we find for the Magneticum ETGs.
This is especially interesting since the evolution found for the total density slope with redshift from these strong lensing observations is so different from that found in the simulations.
That indicates that there might be some selection bias towards galaxies with flatter slopes at higher redshift in the observations.

In the right panel of Fig.~\ref{fig:gamma_sigma_fdm}, $\Sigma_*$ versus $f_\mathrm{DM}$ is shown compared to the Coma cluster ETGs from \citet{thomas:2007MNRAS.382..657T} (open light green circles) and the strong lensing results from \citet{barnabe:2011MNRAS.415.2215B} (open green diamonds).
In this case, the agreement between strong lensing observations and simulations is rather poor for both simulations.
However, the Coma cluster ETGs show a good agreement with the fit to the Magneticum ETGs, and therefore support the evolutionary trend found from the simulations.
\begin{figure}
  \begin{center}
    \includegraphics[width=\columnwidth]{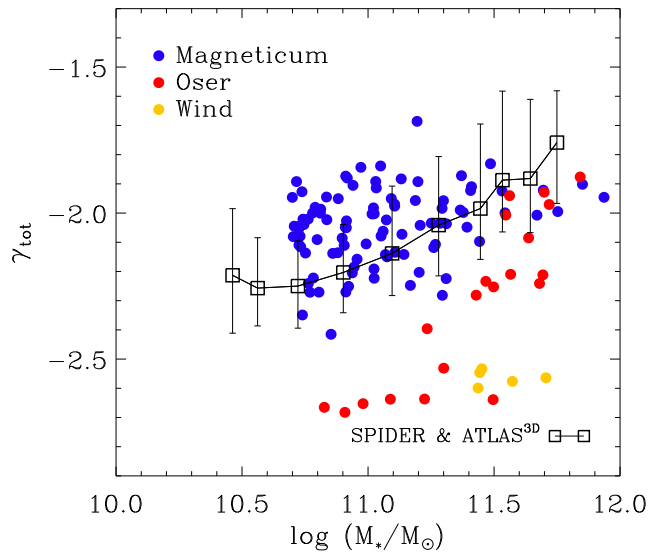}
  \caption{Total mass density slopes $\gamma_\mathrm{tot}$ versus stellar mass $M_*$ at $z=0$, but with observations from the SPIDER and ATLAS$^\mathrm{3D}$ surveys from \citet{tortora:2014a} included as black open squares for comparison.
  Simulations are shown as blue (Magneticum ETGs), red (Oser ETGs) and yellow (Wind ETGs) circles.
}
  {\label{fig:gamma_mass_z0_detail}}
  \end{center}
\end{figure}

One last but very interesting comparison between observations and simulations is shown in Fig.~\ref{fig:gamma_mass_z0_detail}.
Here we show the slopes of the total density profiles $\gamma_\mathrm{tot}$ versus the stellar mass $M_*$ at $z=0$ for all our simulations, and compare the result with observations from \citet{tortora:2014a}, shown as black open squares.
While the Oser ETGs show a strong trend between the slope of the density profile and the stellar mass, the Magneticum ETGs only show a much weaker trend.
As can be seen, the observations are in excellent agreement with the Magneticum ETGs, which clearly demonstrates that the AGN feedback is an essential ingredient for properly modelling the higher-mass ETGs.
This is also in agreement with the results from the recent work by \citet{forbes:2016} from the SLUGGS survey, who also find only a very weak trend between the stellar mass and the total density slopes of their ETGs.

\section{Summary and Discussion}\label{sec:discussevol}
We analyse ETGs selected from three different simulation sets over a redshift range from $0<z<2$ to study the evolution of the central dark matter fractions and the slopes of the total density profiles.
The first set of ETGs are selected from the hydrodynamic cosmological box simulation Magneticum, which include both AGN and stellar feedback as well as metal cooling.
For this ensemble of galaxies, we identify at each timestep all galaxies more massive than $M_\mathrm{*}=5\times10^{10}M_\odot$, and select those which classify as ETGs following \citet{teklu:2015paper}.
The ETGs at high redshift can, but do not necessarily need to, be the progenitors of ETGs at lower redshifts.
At $z=0$, we identify 96 out of 269 galaxies as ETGs.

The second set of 20 ETGs is selected from the sample of cosmological zoom simulations presented by \citet{oser:2010ApJ...725.2312O}.
This simulation sample includes neither AGN nor stellar feedback, and cooling is based on primordial abundances only.
At each timesteps analysed in this work, the Oser ETGs are the progenitors of the 20 ETGs selected at $z=0$.
At $z=0$, we include a third set of simulations, namely 5 ETGs selected from a sample of galaxy zoom-in simulations which include a momentum-driven stellar feedback and metal evolution model but no feedback from AGNs. 
As that simulation sample did not form ETGs at higher redshifts, we only include them at $z=0$.
All samples, however, are limited to central halo galaxies, excluding ETGs that are substructures.

We study the evolution of the mass-size relation with redshift and find a generally good agreement with the observations, at all redshifts.
At low redshifts, our relations found from the simulated ETGs at the high mass end are in good agreement with the mass-size relation found for the SDSS sample by \citet{shen:2003}, while at the lower mass range the simulations more closely resemble the results from the GAMA survey \citep{baldry:2012}, which is focussed especially on the low mass end of the galaxy mass function.
Both simulated samples of ETGs show a similar scatter around the relation as well as similar evolution trends for the high redshift ETGs to have smaller sizes, but the Magneticum ETGs are generally less massive than the Oser ETGs of the same size at all redshifts.
This is due to the fact that the Oser ETGs do not include a central black hole and the associated AGN feedback, which, in case of the Magneticum ETGs, efficiently prevents the overcooling problem by heating part of the cold gas in the centers of the galaxies during their evolution.
In comparison, the Wind ETGs show smaller sizes at comparable masses at the high mass end as the Oser ETGs, which is due to the lack of AGN feedback and an even stronger overproduction of stars in the center due to metal cooling and late re-accretion of previously blown-out gas, overall increasing the in-situ fraction and lowering the fraction of accreted stellar material (see \citealt{hirschmann:2013}).

Additionally, we analyse the evolution of the central dark matter fraction of ETGs.
We find that the central dark matter fractions generally decrease with redshift for all our ETG samples, in good agreement with recent observational results by \citet{tortora:2014b}.
While there is only a weak correlation with the masses of the galaxies, the central dark matter fractions clearly correlate with the sizes of the ETGs.
Furthermore, we find a strong anti-correlation between the central dark matter fraction and the in-situ fraction of ETGs, i.e., ETGs with high central dark matter fractions have only very little stars formed in situ.
This is a natural consequence of the late growth through (dry) merger events \citep{hilz:2012MNRAS.425.3119H,hilz:2013MNRAS.429.2924H}.

As shown by \citet{remus:2013}, the in-situ fractions also correlate with the slopes $\gamma_\mathrm{tot}$ of the total (stellar plus dark matter) radial density profiles.
These slopes are on average close to $\gamma_\mathrm{tot}\approx-2$, indicating that the ETGs are close to isothermal.
However, they show a large scatter and can be as steep as $\gamma_\mathrm{tot}\approx-3$.
Interestingly, we find clear correlations between the steepness of the total density slopes and other quantities studied in this work:
\begin{list}{}{\usecounter{enumi}\itemindent 0em\listparindent 2em\leftmargin 2em\labelwidth 2em\def\makelabel#1{\hss\llap{$\bullet$}}}
\item A tight co-evolution exists between the central dark matter fractions and the slopes of the total density profiles: Galaxies with larger central dark matter fractions have flatter slopes.
This co-evolution can be described as $\gamma_\mathrm{tot} = Af_\mathrm{DM} + B$ and holds for all systems at all redshifts, independent of the different feedback models.
\item The values for $A$ and $B$ change distinctively with the assumed feedback model, and thus this relation can be used as a test for feedback models.
\item A similar correlation exists between $\gamma_\mathrm{tot}$ and the stellar mass surface density $\Sigma_*$, in that compact ETGs tend to have steeper total density slopes than their more extended counterparts. 
\end{list}
This is in agreement with observations from strong lensing \citep[e.g.,][]{sonnenfeld:2013} and dynamical modelling \citep{tortora:2014a}.
As expected, the model with weak stellar feedback and, in particular, feedback from black holes is in better agreement with observations.

All simulations, independent of the assumed feedback model, predict steeper total density slopes and lower dark matter fractions at higher redshifts.
While the latter is in agreement with the observed trends, the former is inconsistent with current lensing observations (\citealt{treu:2004,auger:2010ApJ...724..511A,ruff:2011ApJ...727...96R,sonnenfeld:2013}) who find no changes in the total density slopes with redshift, or, if any, a tendency towards flatter slopes at higher redshifts.
We find this discrepancy between observations and simulations to be a result of the method used observationally to determine the density slopes.
The non-evolution of the total density slopes with redshift can be reproduced for the simulated ETGs by applying the same observational method.
Thus, we conclude that the apparent disagreement between the redshift evolution of the total density slopes from observations and simulations is not real.

In summary, we find clear indications from all sets of simulations for the two-phase evolution scenario for central ETGs:
At high redshift, gas dominates the mass growth of (early-type) galaxies, thus many stars are formed in situ and only few are accreted.
The gas dissipates its energy and sinks to the center of the potential well where it forms the stars in a compact central structure, thus the dark matter fractions are small and the total density slopes are steeper.
At lower redshifts, (dry) merger events of all mass ratios start to dominate the mass growth of the galaxies, leading to an enhanced growth in size compared to the growth in mass, as mass is mostly added to the outskirts (apart from the rare major merger events which actually mix the whole galaxy).
This leads to a growth of the central dark matter fraction, a flattening of the total density slopes and a decrease of the in-situ fraction of stars.
\textit{Indeed, we find a very close correlation between the central dark matter fractions and the slopes of the total radial density profiles of ETGs at all redshifts.
Thus, we conclude that the central dark matter fractions and the slopes of the total radial density profiles of ETGs are good indicators for the amount of dry merging events a galaxy has undergone.}

Generally, we find that the AGN feedback leads to less compact ETGs with higher central dark matter fractions and flatter slopes of the total density profiles.
Therefore, the co-evolution of both quantities can be used as a test for feedback models, as the slopes of this correlation, albeit constant in time, differ depending on the included feedback mechanisms.

\section*{Acknowledgments}
We thank St\'ephane Courteau and Crescenzo Tortora for helpful discussions and comments, and Alessandro Sonnenfeld for providing his code used for determining the density slopes observationally. We also thank the anonymous referee for helpful comments.
The Magneticum Pathfinder simulations were partially performed at the Leibniz-Rechenzentrum with CPU time assigned to the Project ``pr86re''.
This work was supported by the DFG Cluster of Excellence ``Origin and Structure of the Universe''.
We are especially grateful for the support by M. Petkova through the Computational Center for Particle and Astrophysics (C2PAP).
TN acknowledges support by the DFG priority program 1573.
MH acknowledges financial support from the European Research Council via an Advanced Grant under grant agreement no. 321323 NEOGAL.
PHJ acknowledges the support of the Academy of Finland grant 1274931.

\bibliographystyle{aa}
\bibliography{bibliography}

\label{lastpage}
\end{document}